# The Gridbus Toolkit for Service Oriented Grid and Utility Computing: An Overview and Status Report

Rajkumar Buyya and Srikumar Venugopal

Grid Computing and Distributed Systems Laboratory
Department of Computer Science and Software Engineering
The University of Melbourne, Australia
*{raj, srikumar}@cs.mu.oz.au*

**Abstract:**
*Grids aim at exploiting synergies that result from cooperation of autonomous distributed entities. The synergies that result from grid cooperation include the sharing, exchange, selection, and aggregation of geographically distributed resources such as computers, data bases, software, and scientific instruments for solving large-scale problems in science, engineering, and commerce. For this cooperation to be sustainable, participants need to have economic incentive. Therefore, "incentive" mechanisms should be considered as one of key design parameters of Grid architectures. In this article, we present an overview and status of an open source Grid toolkit, called Gridbus, whose architecture is fundamentally driven by the requirements of Grid economy. Gridbus technologies provide services for both computational and data grids that power the emerging eScience and eBusiness applications.*

## 1 Introduction

Grid computing [2] has emerged as a new paradigm for next-generation computing. It supports the creation of virtual organizations and enterprises that enable the sharing, exchange, selection, and aggregation of geographically distributed heterogeneous resources for solving large-scale problems in science, engineering, and commerce. The Grid community has embraced the integration of commodity Web services and Grid technologies, which has led to the development of Grid services [6]. The widespread interest in grid computing from commercial organisations in recent times is pushing it towards the mainstream, thus enhancing Grid services to become valuable economic commodities.

In spite of a number of advances in Grid computing, resource management and scheduling in such environments continues to be a challenging and complex undertaking. One of the problems is dealing with geographically distributed resources owned by different organizations with different usage policies, cost models and varying load and availability patterns. The grid service providers (resource owners) and grid service consumers (resource users) have different goals, objectives, strategies, and requirements. To address these resource management challenges, distributed computational economy has been recognized as an effective metaphor for the management of Grid resources [4, 5] as it: (1) enables the regulation of supply and demand for resources, (2) provides economic incentive for grid service providers, and (3) motivates the grid service consumers to trade-off between deadline, budget, and the required level of quality-of-service. These features are essential for commodity Grid services.

The idea of a computational economy helps in creating a service-oriented computing architecture where service providers offer paid services associated with a particular application and users, based on their requirements, would optimize by selecting the services they require and can afford within their budget. To realize this scenario, the Gridbus project [7] is actively pursuing research in the design and development of open source cluster and grid middleware technologies for utility and service-oriented computing. Gridbus emphasizes the end-to-end quality of services driven by computational economy at various levels - clusters, peer-to-peer (P2P) networks, and the Grid - for the management of distributed computational, data, and application services.

At the cluster level, the Libra scheduler has been developed to support economy-driven cluster resource management. Libra is used within a single administrative domain for distributing computational tasks among resources that belong to a cluster. At the P2P network level, the CPM (compute-power-market) infrastructure is being developed through the JXTA community. At the Grid level, various tools are being developed to support a quality-of-service (QoS) - based management of resources and scheduling of applications. To enable performance evaluation, a Grid simulation toolkit called GridSim has been developed. GridSim supports the modeling and simulation of application scheduling on simulated Grid resources. Finally, to support the accounting of resource or service usage and enable sustainable resource sharing across virtual organizations, we have developed Grid Accounting Services infrastructure.

## 2   Gridbus System Vision and Architecture

Scientific discoveries and business decisions today are increasingly driven by analysis of data. Some of the target data-intensive applications that motivates our work include high-energy physics, molecular docking for drug discovery, and neuroscience. Drug designers conduct computationally intensive molecular docking technique to screen/analyse large-scale, distributed chemical databases to identify macromolecules that potentially serve as drug candidates. Businesses use various data mining techniques in decision support systems that analyse customer transaction records. In such data-intensive environments, there is a huge load on precious resources such as network bandwidth, computational and storage resources. Grid economy can be used to regulate the usage of these resources by using differential pricing strategies that provide users with incentives to trade-off lower costs for more relaxed timeframes and to use resources at off-peak hours.

The Gridbus Project is investigating solutions for enabling such value-based interactions within a data-intensive computing environment. Figure 1 depicts a distributed data-oriented application scenario within which the Gridbus Project components have been deployed in conjecture with other middleware and hardware technologies.

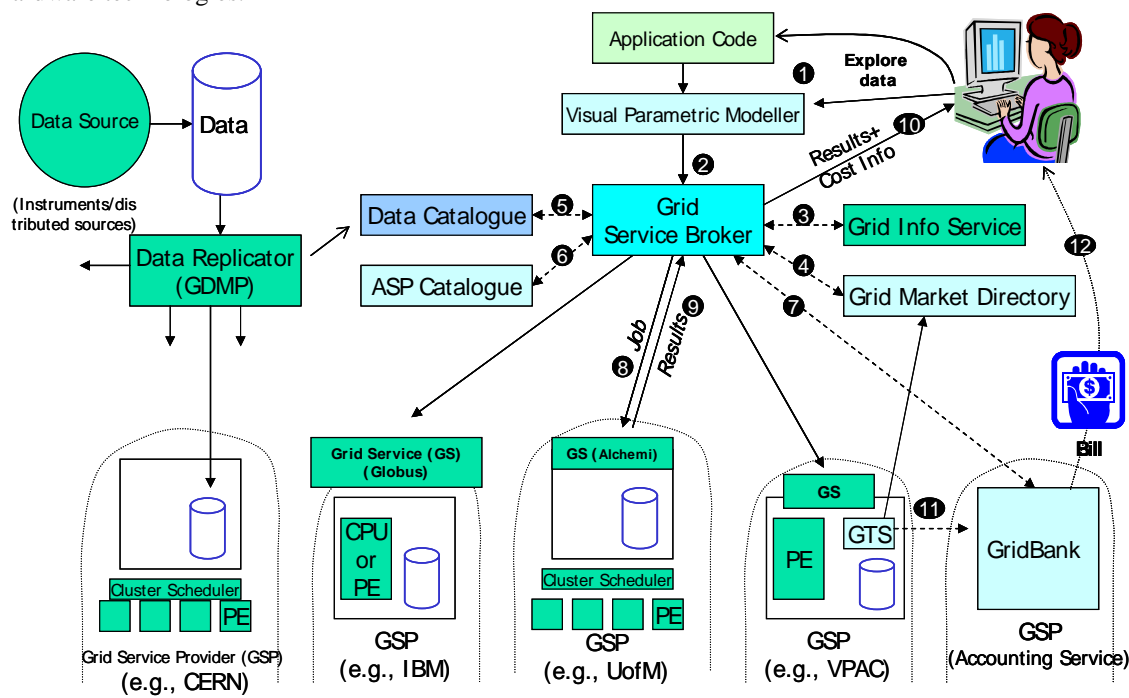

**Figure 1: A Utility Grid Architecture with Grid Economy.**

The steps involved in analysing distributed data are as follows.  The application code is the legacy application has to be executed on a grid.  The users compose their application as a distributed application (e.g., parameter sweep) using visual application development tools (Step 1). The parameter-sweep model of creating several independent jobs is well suited for grid computing environments wherein challenges such as load volatility, high network latencies and high probability of failure of individual nodes make it difficult to adopt a programming approach which favours tightly coupled systems. Accordingly, this has been termed as a "killer application" for the Grid [3]. Visual tools allow rapid composition of applications for grids while taking away the associated complexity.

The user's analysis and quality-of-service requirements are submitted to the Grid resource broker (Step

2). The Grid resource broker performs resource discovery based on user-defined characteristics, including price, using the Grid information service and the Grid Market Directory (Steps 3&4). The broker identifies the list of data sources or replicas and selects the optimal ones (Step 5). The broker also identifies the list of computational resources that provides the required application services using the Application Service Provider (ASP) catalogue (Step 6). The broker ensures that the user has the necessary credit or authorized share to utilise resources (Step 7). The broker scheduler maps and deploys data analysis jobs on resources that meet user quality-of-service requirements (Step 8). The broker agent on a resource executes the job and returns results (Step 9). The broker collects the results and passes them to the user (Step 10).The metering system charges the user by passing the resource usage information to the accounting system (Step 11). The accounting system reports resource share allocation or credit utilisation to the user (Step 12).

In the following sections, we briefly discuss some of the Gridbus technologies shown in Figure 1.

## 3  Gridbus Technologies

The Gridbus Project is engaged in the design and development of grid middleware technologies to support eScience and eBusiness applications. These include visual Grid application development tools for rapid creation of distributed applications, competitive economy-based Grid scheduler, cooperative economy-based cluster scheduler, Web-services based Grid market directory (GMD), Grid accounting services, Gridscape for creation of dynamic and interactive testbed portals, G-monitor portal for web-based management of Grid applications execution, and the widely used GridSim toolkit for performance evaluation. Recently, the Gridbus Project has developed a Windows/.NET-based desktop clustering software and Grid job web services to support the integration of both Windows and Unix-class resources for Grid computing. A layered architecture for realisation of low-level and high-level Grid technologies is shown in Figure 2. Some of the Gridbus technologies discussed below have been developed by making use of Web Services technologies and services provided by low-level Grid middleware, particularly Globus Toolkit [1] and Alchemi. A summary and status of various Gridbus technologies is listed in Table 1.

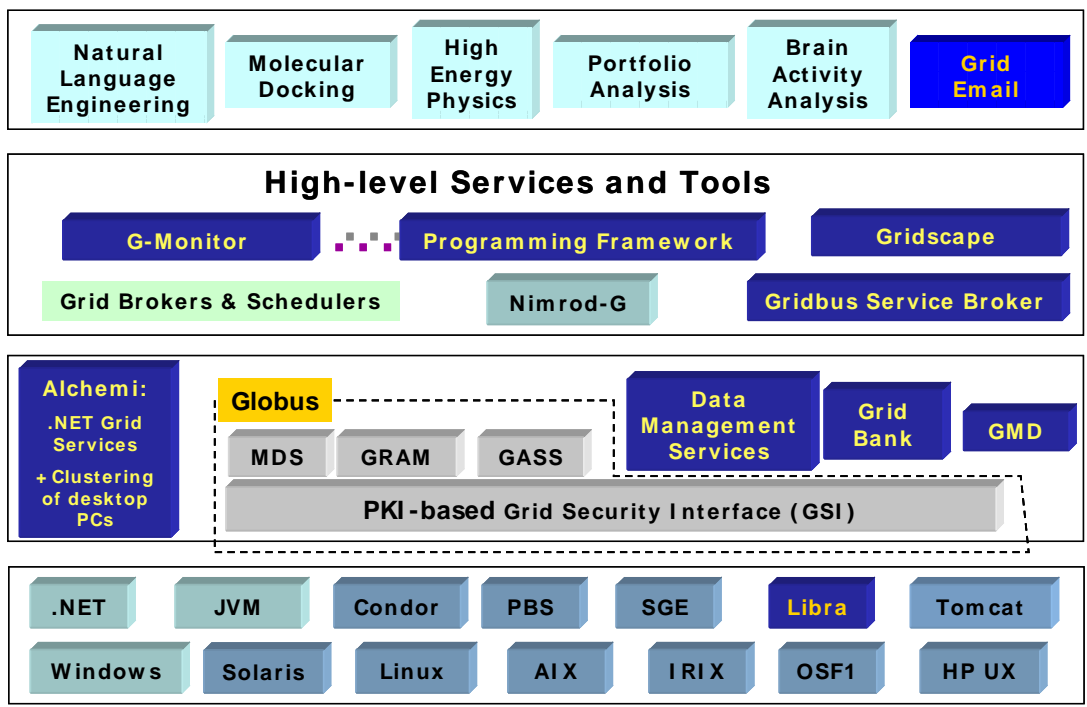

**Figure 2: A Grid technology stack (items in the dark blue box are pursued by the Gridbus project).**

### 3.1  Visual Parameter Sweep Application Composer

The Gridbus project developed a Java based IDE, called Visual Parametric Modeler (VPM), for rapid

creation of parameter sweep (data parallel/SPMD) applications [14]. VPM allows users to parameterize the input data files to transform static values to variable parameters, and to create a script that defines parameters and tasks. VPM also allows the rapid creation and manipulation of the parameters. While being flexible, it is also simple enough for a non-expert to create a parameter script, known as a plan file, within minutes. The parameter sweep applications composed using VPM can be deployed on global Grids using the Gridbus resource broker.

| Gridbus Component | Description | Current Status |
|---|---|---|
| Visual Parametric Modeller | A graphical environment for application parameterisation. | Ver.1.0 with visual parameterisation of data files. |
| G-Monitor | A web portal to manage execution of applications on Grids using remote brokers. | Ver.2.0 with support for access from handhelds. |
| Grid Service Broker | An economy-based Data Grid broker for scheduling distributed data oriented applications across Windows and Unix-variant Grid resources. | Ver.1.0 with dynamic parameters whose value is determined at runtime. |
| Grid Market Directory | A directory for publication of Grid Service Provides and their services. | Ver.1.0 with web services based query interface. |
| Grid Bank | A grid accounting, authentication, and payment management infrastructure. | Ver.1.0 with web services interface. |
| Gridscape | A tool for the creation of interactive and dynamic Grid testbed web portals. | Ver.1.2 implemented as Web application within Tomcat. |
| Alchemi | A .NET-based desktop Grid framework. | Ver.0.65 with interface for user-level scheduling. |
| Libra | An economy based scheduler for clusters. | Ver.1.0 implemented as PBS plug-in. |
| GridSim | A toolkit for modelling and simulation of global Grids. | Ver 3.0 with advance resource reservation. |

**Table 1: Gridbus technologies and their status as of April 2004.**

## 3.2   G-Monitor

G-Monitor [15] is a web portal for initiating, monitoring and steering application execution on global grids. It uses services provided by Grid Service Brokers such as Nimrod-G and Gridbus Broker to deploy applications. It allows users to manage their Grid credentials and provides secure access to remote hosts running brokers. The users can either upload applications and data at runtime or select from those already available on the broker host. When a user makes a request to start execution of his application, G-Monitor authenticates him to the broker host and creates a Grid proxy on it valid for the specified duration. At the end of the execution, the users can download the collected results from the broker host to their workstation.

G-Monitor provides an easy to use interface for the end-user to monitor and control jobs running within the Grid environment. It depicts the execution progress via graphs which can be viewed by the user at anytime. G-Monitor is also scalable, and can therefore handle thousands of nodes and jobs running in a Grid environment. It has been enhanced to identify different access devices such as Pocket PCs and desktop PCs and provide appropriate user interface.

## 3.3   Gridbus Grid Service Broker

The Gridbus Broker [20] makes scheduling decisions on where to place the jobs on the Grid depending on the computational resources characteristics (such as availability, capability, and cost), the users's quality-of-service requirements such as the deadline and budget, and the proximity of the required data or its replicas to the computational resources. Catalogues of replicated data describe the size of the file, the location of the file, the date it was produced, the number of events and other such attributes. Given a job and the input file(s) it requires, the broker looks up the Replica Catalogue at the local site to locate the sites where the required input file is and its size. Then it takes into account various other factors such as the cost, the computing power available at the site, the network bandwidth, the resource reputation, and the account information to make a decision on where to dispatch the job. The broker identifies resource service prices by querying the Grid market directory (GMD).

If an application needs to access remote databases, we provide transparent data access mechanisms and a catalogue that supports logical mapping of data files to a distributed storage devices. The broker performs discovery and online extraction of data-sets from the closest data sources and then farms out analysis jobs to optimal resources. The broker evaluates whether to process jobs on a resource where the data is available by moving the application code, move data to a resource where the application is available, or move both of them to a suitable computing resource.

### 3.4 Grid Market Directory (GMD)

The Grid Market Directory [9] serves as a registry for high-level service publication and discovery in virtual organizations. It enables service providers to publish the services which they provide along with the costs associated with those services. Consumers browse GMD to find services that meet their requirements.

GMD is built over standard Web service technologies such as SOAP (Simple Object Access Protocol) and XML. Therefore, it can be queried by other programs. To provide with an additional layer of transparency, a client API (Application Programming Interface) has been provided that could be used by programs to query the GMD without the developers having to concern themselves with SOAP details. The Gridbus scheduler interacts with the GMD to discover the testbed resources and their high-level attributes such as access price.

### 3.5 GridBank

GridBank (GB) [11] is a secure Grid-wide accounting and (micro) payment handling system. It maintains the users' (consumers and providers) accounts and resource usage records in a database. GridBank supports protocols that enable its interaction with the resource brokers of Grid Service Consumers (GSCs) and the resource traders of Grid Service Providers (GSPs). It has been primarily designed to provide services for enabling a Grid computing economy; however, we envision its usage in e-commerce applications as well. The GridBank services can be used in both co-operative and competitive distributed computing environments.

### 3.6 Gridscape

As more and more people are beginning to embrace grid computing and thus are seeing the need to set up their own grids and grid testbeds, there is a need to have some means to enable them to view and monitor the status of the resources in these testbeds (eg. Web based Grid portal). Generally developers invest a substantial amount of time and effort developing custom monitoring software. Gridscape [18] has been developed to overcome this limitation. It is a tool that enables the rapid creation of interactive and dynamic testbed portals without any programming effort. Gridscape primarily aims to provide a solution for those users who need to be able to create a grid testbed portal but do not necessarily have the time or resources to build a system of their own from scratch..

Gridscape consists of two key individual components - a web application and a related administrating tool-- that are implemented in Java by following MVC (Model-View-Controller) based, Model-2 type architecture. It has been designed to meet the following aims:

- Allow for the rapid creation of Grid testbed portals;
- Allow for simple portal management and administration;
- Provide an interactive and dynamic portal;
- Provide a clear and user-friendly overall view of Grid testbed resources; and
- Have a flexible design and implementation such that core components can be leveraged, it provides a high level of portability, and a high level of accessibility (from the browsers perspective).

### 3.7 Alchemi

Software to enable grid computing has been primarily written for Unix-class operating systems, thus severely limiting the ability to effectively utilize the computing resources of the vast majority of desktop computers i.e. those running variants of the Microsoft Windows operating system. Addressing Windows-based grid computing is particularly important from the software industry's viewpoint where interest in grids is emerging rapidly. Microsoft's .NET Framework has become near-ubiquitous for implementing commercial distributed systems for Windows-based platforms, positioning it as the ideal platform for grid computing in this context.

Alchemi [19] is a .NET-based grid computing framework that provides the runtime machinery and programming environment required to construct desktop grids and develop grid applications. It allows flexible application composition by supporting an object-oriented grid application programming model in addition to a grid job model. Cross-platform support is provided via a web services interface and a flexible execution model that supports dedicated and non-dedicated (voluntary) execution by grid nodes.

### 3.8 Libra

Clusters of computers have emerged as mainstream parallel and distributed platforms for high-performance, high-throughput and high-availability computing. To enable effective resource management on clusters, numerous cluster management systems and schedulers have been designed. However, their focus has essentially been on maximizing CPU performance, but not on improving the value of utility delivered to the user and quality of services. The Gridbus Project developed a new computational economy driven scheduling system called Libra [22], which has been designed to support allocation of resources based on the users' quality of service (QoS) requirements. It is intended to work as an add-on to the existing queuing and resource management system. The first version has been implemented as a plugin scheduler to the PBS (Portable Batch System) system. The scheduler offers market-based economy driven service for managing batch jobs on clusters by scheduling CPU time according to user-perceived value (utility), determined by their budget and deadline rather than system performance considerations. The Libra's scheduling algorithm shows that the deadline and budget based proportional resource allocation strategy improves the utility of the system and user satisfaction as compared to system-centric scheduling strategies. We believe that such a feature of Libra helps in enforcing resource allocation based on service-level agreements when cluster services are offered as a utility on the Grid.

### 3.9 GridSim

The GridSim toolkit [12] supports modeling and simulation of a wide range of heterogeneous resources: single- or multiprocessors, shared and distributed memory machines, such as PCs, workstations, SMPs, and clusters with different capabilities and configurations. GridSim can be used for modeling and simulation of application scheduling on various classes of parallel and distributed computing systems, such as clusters, grids, and P2P networks. The GridSim resource entities are being extended to support advanced reservation of resources and user-level setting of background load on simulated resources based on trace data.

The GridSim toolkit provides facilities for the modeling and simulation of resources and network connectivity with different capabilities, configurations, and domains. It supports primitives for application composition, information services for resource discovery, and interfaces for assigning application tasks to resources and managing their execution. It also provides visual modeler interface [13] for creating users and resources. These features can be used to simulate parallel and distributed scheduling systems such as resource brokers or Grid schedulers for evaluating performance of scheduling algorithms or heuristics. The GridSim Toolkit has been used to create a resource broker that simulates Nimrod-G for the design and evaluation of deadline and budget constrained scheduling algorithms with cost and time optimizations. It is also used to simulate a market-based cluster scheduling system (Libra) in a cooperative economy environment.

Recently GridSim has been extended to support the simulation of Advanced Reservation of resources. One of our collaborators has developed modules to support the simulation of Data Grid environment within GridSim.

# 4   Sample eScience Applications

The *distribution* of knowledge (by scientists) and data sources (advanced scientific instruments), and the *need* of large-scale computational resources for analyzing massive scientific data are two major problems commonly observed in scientific disciplines. Two scientific disciplines of this nature are brain science and high-energy physics.

The Gridbus Project has been actively involved in extending its technologies to Grid enable real-world applications in collaboration with various researchers around the world (see Table 2). In addition, we have extended the notion of Grid economy to develop an attention economy based eMail communication system, called GridMail [17]. The idea is to regulate electronic communication between people through attention economy, which can eventually kill spam problem that we are facing in the cyber world. A brief discussion on how two of these applications have been Grid-enabled and deployed on global Grids is given below.

| Application area | Contact |
|---|---|
| Molecular Docking [8] (Drug Discovery) | Dr. Kim Branson, Ludwig Institute for Cancer Research, Melbourne |
| Neuroscience (Brain Activity Analysis) | Dr. Susumu Date, School of Medicine, Osaka University, Japan |
| High Energy Physics | Dr. Martin Sevior, School of Physics, University of Melbourne |
| Finance (Portfolio analysis) | Dr. Rafael Moreno Vozmediano, Complutense University of Madrid, Spain |
| Natural Language Engineering | Dr. Steven Bird, Dept. of Computer Science, University of Melbourne |
| Astrophysics | Dr. David Barnes, School of Physics, University of Melbourne |
| Australian Earth and Ocean Network (AEON) | Dr. Dietmar Müller, Institute of Marine Sciences, University of Sydney |
| Molecular and Materials Structure Network | Dr. Peter Turner, School of Chemistry, University of Sydney |
| Structural Engineering | Dr. Priyan Mendis, Department of Civil and Environmental Engineering, University of Melbourne |

**Table 2: Key application drivers for the Gridbus project and collaborators.**

## 4.1   Neuroscience and Virtual Instrumentation

The analysis of brain activity data gathered from the MEG (Magnetoencephalography) instrument is an important research topic in medical science since it helps doctors in identifying symptoms of diseases. The data needs to be analyzed exhaustively to efficiently diagnose and analyze brain functions and requires access to large-scale computational resources. We collaborated with Osaka University, Japan and worked with them to design and develop MEG data analysis system by leveraging Grid technologies, primarily Nimrod-G, Gridbus, and Globus. The neuroscience (brain activity analysis) application has been formulated as parameter-sweep application and demonstrated it's potential as an eScience application by deploying it on the World-Wide Grid testbed [10].

## 4.2   The Belle High Energy Physics Experiment

The Belle experiment located at the KEK Particle Accelerator, Tsukuba, Japan, is probing Charge-Parity (CP) violation in the Standard Model (SM) of Physics. It involves a collaboration of 400 researchers across 50 institutes from 10 countries and provides the state of the art instrument to detect and reconstruct the production and decay of the mesons produced at the KEK B-factory. The increasing efficiencies of the KEKB accelerator have led to an increase in the rate of data production from the Belle experiment. The current experiment and simulation data set is nearly 10 Terabytes (TB) in size. Also, this data is distributed globally and the locations of the data repositories are provided in the Replica Catalog. Hence, this project will greatly benefit from application of data grid techniques.

We have used the Gridbus broker to grid-enable the Belle Analysis Software Framework (BASF), the main application for Belle data analysis [20]. During execution, the broker mediated access to distributed resources by (a) discovering suitable data sources for a given analysis scenario, (b) suitable computational resources, (c) optimally mapping analysis jobs to resources, (d) deploying and monitoring job execution on selected resources, (e) accessing data from local or remote data source during job execution and (f) collating and presenting results. During execution, the broker was able to reduce the cost of network transfer by allocating jobs to the best available computational resources that were closest to the sources of the input data for that job which also reduced the data transfer time. Currently, we are extending this research by introducing deadline and budget as constraints within this distributed data-oriented environment.

## 5   Gridbus Deployment in the Global Data-Intensive Grid Collaboration

The Gridbus Project has led the establishment a world-wide collaboration, called the Global Data-Intensive Grid Collaboration [16], with of aim of creating a virtual organisation to demonstrate a large number of distributed data-intensive computing applications by harnessing geographically distributed resources. The collaboration was nominated as one of finalists for HPC Challenge event organized as a part of the IEEE/ACM Supercomputing Conference (SC 2003) held at Phoenix, Arizona, USA from Nov. 15-21, 2003. The collaboration has put together a World-Wide Grid (WWG) that contains over 200 Grid nodes (PCs, workstations, clusters, supercomputers, databases, and applications) contributed by organizations and volunteers based in Australia, Asia, Europe, North America, and South America. A snapshot of status of WWG resources as visualized using Gridscape is shown in Figure 3. The testbed supported included resources running Unix-variant or Windows operating systems.

Some of the vital statistics of the testbed are as follows:
- Grid Nodes: 218 distributed across 62 organizations from 21 countries around the world.
    - Laptops, desktop PCs, WS, SMPs, Clusters, supercomputers
- CPU Architecture:
    - Intel x86, IA64, AMD, PowerPC, Alpha, MIPS
- Operating Systems:
    - Windows or Unix-variants – Linux, Solaris, AIX, OSF, Irix, HP-UX
- Intranode Network:
    - Ethernet, Fast Ethernet, Gigabit, Myrinet, QsNet, PARAMNet
- Internet/Wide Area Networks
    - GrangeNet, AARNet, ERNet, APAN, TransPAC, and so on.
- Grid Middleware:
    - Alchemi for access to Windows nodes and Globus for Unix-variants.
    - The Gridbus Service Broker for both Windows and Unix-variants resources.
    - Nimrod-G for accessing Unix-variants resources for running GAMESS application.
    - Other Gridbus technologies such as Gridscape, G-Monitors as described above.

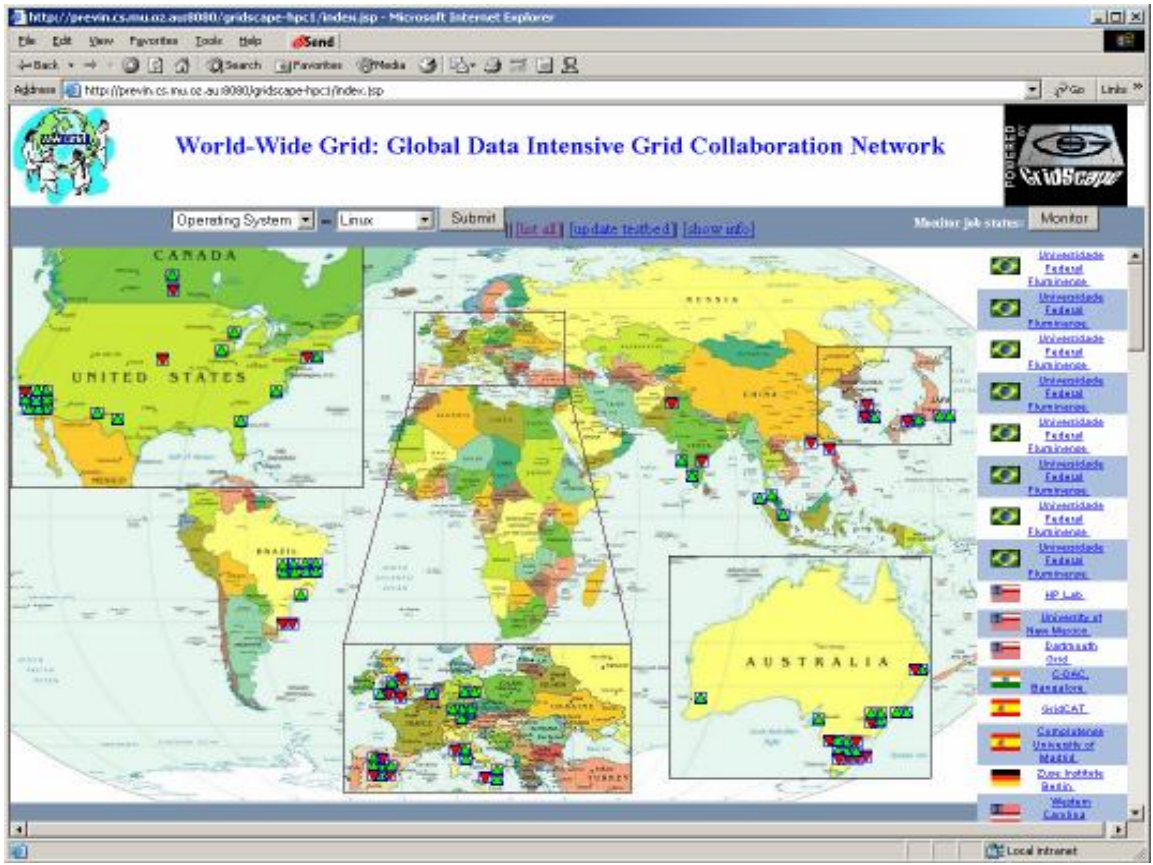

**Figure 3: World Wide Grid (WWG).**

The contributors with resources running Unix-variant OSes have provided Grid access through Globus whereas those running Windows and .NET have provided Grid access through Alchemi middleware. The collaboration demonstrated on demand deployment of various data-intensive computing applications from natural language processing and particle physics to portfolio analysis on the WWG using the Grid Service Broker (GSB). The broker was able to simultaneously utilise both Alchemi and Globus-based resources and deploy appropriate application codes at runtime. Nimrod-G broker developed by Monash University was used in demonstrating a quantum chemistry application (GAMESS) on the testbed resources [21]. A summary of applications execution statistics during the HPC Challenge demonstration is shown in Table 3. The results of each application are interpreted using application specific graphical visualisation tools.

| Application | Data Size | Processing Time | Nodes |
|---|---|---|---|
| Belle Analysis (HEP) | 300 MB input (100 jobs – 3MB each) | 30 min. | Australia, Japan |
| Financial Portfolio Analysis | 50 MB output (50 jobs – 1MB each) | 20 min. | Global |
| Newswire Indexing | 80 MB input (12 jobs – 7MB each job) | 20 min. | GrangeNet, Australia |
| GAMESS Quantum chemistry application | 4KB for each job. Total output: 860MB compressed | Each job took 5-78 minutes. Total 15 hours | Global (130 nodes, 15 sites) |

**Table 3: A summary of applications execution on the WWG testbed demonstrated at SC2003.**

# 6  Conclusion

We have presented an overview of the Gridbus toolkit for service oriented grid and utility computing based on computational economy. The Gridbus project is actively pursuing the design and development of next-generation computing systems and fundamental Grid technologies and algorithms driven by Grid economy for data and utility Grid applications.

The Gridbus Project is continuously enhancing and building on the various Grid technologies presented in this article. The project is also actively investing and developing new Grid technologies such as Grid Exchange that enable the creation of a Stock Exchange like Grid computing environment. For detailed and up-to-date information Gridbus technologies and new initiatives, please visit the project web site http://www.gridbus.org.

## Acknowledgement

Some of the Gridbus technologies are being developed by extending the early work carried out by Rajkumar Buyya with David Abramson, Jon Giddy, Manzur Murshed, and Kim Branson. The continued contributions of Jia Yu, Martin Placek, Alexander Barmouta, Anthony Sulistio, Chee Shin Yeo, Rajiv Ranjan, Ding Choon Hong, Hussein Gibbins, and Akshay Luther to the Gridbus Project are highly appreciated. We thank all members of the Global Data Intensive Grid Collaboration and all our Grid collaborators for their contribution.

The Gridbus Project has been sponsored by the University of Melbourne, Sun Microsystems, Victorian Partnership for Advanced Computing (VPAC), Australian Research Council (ARC), International Business Machine (IBM), Singapore Computer Systems (SCS), and Storage Technology Corporation.